\documentstyle[12pt,psfig]{article}
\def\deg{\ifmmode^{\circ}\;\else$^{\circ}\;$\fi}

\begin{document}
\title{Single Dish Polarization Calibration}
\author{Simon Johnston}
\date{}
\maketitle
{\center
School of Physics, University of Sydney, NSW 2006, Australia\\[3mm]}

\begin{abstract}
Using the formalism of Hamaker et al. (1996), I derive a method for
the polarization calibration of observations made with a single
radio telescope. This method is particularly appropriate for observations
of pulsars, where the sign and magnitude of the circular polarization
are useful for understanding the emission processes at work.
I apply the method to observations of PSR J1359--6038
made using the multibeam receiver on the Parkes radio telescope.
\end{abstract}

\section{Introduction}
The polarization properties of astronomical sources are interesting
for a number of reasons. They can be used, for example, in determining
the physical processes at work; thermal sources are not polarized,
whereas non-thermal sources may have a significant degree of polarization.
The rotation measure of a source can be used to determine magnetic
field strengths; for example pulsar rotation measures are used to derive
magnetic maps of the Galaxy.
However, the degree of polarization in astronomical sources is often 
small, particularly circular polarization, and so careful calibration is 
necessary before the results can be interpreted.

Recent papers on polarization calibration of radio telescopes
include those by Hamaker et al. (1996) and Britton (2000).
The Hamaker et al. (1996) paper lays down the mathematical 
foundations which are subsequently used by Sault et al. (1996) to construct
a method for instrumental calibration of a radio interferometer.
Their method is routinely applied to data from the Australia Telescope
Compact Array and incorporated in the software package {\sc miriad}.
In this paper, I will use the approach of Hamaker et al. (1996) to
develop a method for calibration of single dish data. This method
is particularly useful for observations of pulsars. In the final section
of the paper I apply the method to observations of the 
pulsar PSR J1359--6038 using the 20-cm multibeam receiver on 
the Parkes radio telescope and discuss the results.

\section{Representation}
Following Hamaker et al. (1996) I describe the propagation of an
electromagnetic wave in the $xyz$ coordinate system by
\begin{equation}
e_2 = <
\left( \begin{array}{c}
e_x \\ e_y
\end{array} \right)
>
\end{equation}
where the axis of propagation is the $z$ axis,
and $e_x$ and $e_y$ are complex.
The coherency properties of the electric field 
can be expressed in $xy$ coordinates by the coherency matrix
(Born \& Wolf 1964)
\begin{equation}
e = <
\left( \begin{array}{c}
e_x e_x^* \\ e_x e_y^* \\ e_y e_x^* \\ e_y e_y^*
\end{array} \right)
>
\end{equation}
where the $*$ denotes the complex conjugate.
The electromagnetic field is converted to electric voltage in the
{\it feed} of a radio telescope which,
in the case of linear feeds, consists of two input probes 
aligned along $x$ and $y$.
A correlator multiplies and averages these voltages to produce 
the voltage vectors
\begin{equation}
v_2 = <
\left( \begin{array}{c}
v_x \\ v_y
\end{array} \right) >
;\,\,\,\,\,
v = <
\left( \begin{array}{c}
v_x v_x^* \\ v_x v_y^* \\ v_y v_x^* \\ v_y v_y^*
\end{array} \right)
>
\end{equation}
Generally, one is interested in the (true) Stokes parameters, $I$, $Q$, $U$ and
$V$ which in combination form the Stokes vector $e^S$.
One can express $e^S$ in terms of the electric field vector, $e$, by
\begin{equation}
e^S = 
\left( \begin{array}{c}
I \\ Q \\ U \\ V
\end{array} \right)
= T e;\,\,\,\,\, T = 
\left( \begin{array}{rrrr}
1&0&0&1 \\ 1&0&0&-1 \\ 0&1&1&0 \\ 0&-i&i&0
\end{array} \right)
\end{equation}

The propagation of the radiation from the astrophysical source of
interest to the output from 
the correlator can be described in the following way
(Hamaker et al. 1996).
Let the effects of Faraday Rotation (both through the interstellar
medium and the ionosphere) and parallactic angle rotation be
combined into the matrix $R$.
We let the matrix $F$ represent the feed response
(including gains, phases and leakage terms).
Hence the voltages at the two probes can be described by 
\begin{equation}
v_2 =  F\,\, R
\end{equation}
and the correlator voltage vector is given by
\begin{equation}
\label{kron}
v = (F \otimes F^*)\,\, (R \otimes R^*)\,\, T^{-1}\,\, e^S 
\end{equation}
where $\otimes$ denotes the Kronecker (or outer) product (Hamaker et al. 1996).
The rotation matrix, $R$, and the feed response matrix, $F$,
are given by Hamaker et al. (1996) as
\begin{equation}
R =
\left( \begin{array}{rr}
{\rm cos}\phi & {\rm -sin}\phi \\ {\rm sin}\phi & {\rm cos}\phi
\end{array} \right)
; \,\,\,\,\,
F =
\left( \begin{array}{rr}
G&B \\ -C&H
\end{array} \right)
\end{equation}
where each of $B$, $C$, $G$ and $H$ are complex terms.
It is important to note that Hamaker et al. (1996) have shown
that such a formalism does not involve any approximations, although often,
as below, the matrices are then expanded only to first order.

In this formalism, however, unlike that of Britton (2000), there is
no direct physical association of the variables $B$, $C$, $G$ and $H$.
$B$ and $C$ can be thought of as `leakage' terms; they describe
the leakage of the opposite polarization into either receptor.
The real parts of $G$ and $H$ relate to the sensitivity of the
two probes and the difference in the imaginary parts of $G$ and $H$ relates
to the phase offset between the two channels (often called `instrumental
phase' for a pair of linear receptors). However, this is not
strictly correct, as $G$ and $H$ also mixed with the leakage terms
(Hamaker et al. 1996).

Expanding equation \ref{kron} and ignoring (for now) the effects of
rotation we thus obtain
\begin{equation}
\left( \begin{array}{c}
v_{xx^*} \\ v_{xy^*} \\ v_{yx^*} \\ v_{yy^*}
\end{array} \right)
= \frac{1}{2}
\left( \begin{array}{rrrr}
GG^* & GB^* & BG^* & BB^* \\
-GC^* & GH^* & -BC^* & BH^* \\
-CG^* & -CB^* & HG^* & HB^* \\
CC^* & -CH^* & -HC^* & HH^*
\end{array} \right)
\left( \begin{array}{rrrr}
1&1&0&0 \\ 0&0&1&i \\ 0&0&1&-i \\ 1&-1&0&0
\end{array} \right)
\left( \begin{array}{c}
I \\ Q \\ U \\ V
\end{array} \right)
\end{equation}
where the scaling factor 1/2 arises from the determinant of $T$.
Multiplying the two 4x4 matrices together we arrive at
\begin{equation}
\label{huge}
\left( \begin{array}{c}
v_{xx^*} \\ v_{xy^*} \\ v_{yx^*} \\ v_{yy^*}
\end{array} \right)
= \frac{1}{2}
\left( \begin{array}{rrrr}
GG^* + BB^* & GG^* - BB^* & GB^* + BG^* & i(GB^* - BG^*) \\
-GC^* + BH^* & -GC^* - BH^* & GH^* - BC^* & i(GH^* + BC^*) \\
-CG^* + HB^* & -CG^* - HB^* & -CB^* + HG^* & -i(CB^* + HG^*) \\
CC^* + HH^* & CC^* - HH^* & -CH^* - HC^* & -i(CH^* - HC^*) 
\end{array} \right)
\left( \begin{array}{c}
I \\ Q \\ U \\ V
\end{array} \right)
\end{equation}

\section{Measured Stokes Parameters}
Using equation \ref{huge} above and defining the measured Stokes parameters
in the case of linear feeds by
\begin{eqnarray}
I_m & = & v_{xx^*} + v_{yy^*} \\
Q_m & = & v_{xx^*} - v_{yy^*} \nonumber \\
U_m & = & v_{xy^*} + v_{yx^*} \nonumber \\
iV_m & = & v_{xy^*} - v_{yx^*} \nonumber  
\end{eqnarray}
we can express the measured Stokes parameters in terms of the true
Stokes parameters as follows:
\begin{eqnarray}
\label{thelot}
I_m & = & \frac{1}{2} I (g_1^2 + g_2^2 + h_1^2 + h_2^2) + 
          \frac{1}{2} Q (g_1^2 + g_2^2 - h_1^2 - h_2^2) + \\
    &   &             U (g_1 b_1 + g_2 b_2 - h_1 c_1 - h_2 c_2) +
                      V (g_1 b_2 - g_2 b_1 + h_1 c_1 - h_2 c_2) \nonumber \\
Q_m & = & \frac{1}{2} I (g_1^2 + g_2^2 - h_1^2 - h_2^2) + 
          \frac{1}{2} Q (g_1^2 + g_2^2 + h_1^2 + h_2^2) + \nonumber \\
    &   &             U (g_1 b_1 + g_2 b_2 + h_1 c_1 + h_2 c_2) +
                      V (g_1 b_2 - g_2 b_1 - h_1 c_2 + h_2 c_1)\nonumber \\
U_m & = &             I (h_1 b_1 + h_2 b_2 - g_1 c_1 - g_2 c_2) +
                      Q (-g_1 c_1 - g_2 c_2 - h_1 b_1 - h_2 b_2) + \nonumber \\
    &   &             U (g_1 h_1 + g_2 h_2) +
                      V (g_1 h_2 - g_2 h_1) \nonumber \\
V_m & = &             I (g_1 c_2 - g_2 c_1 + h_1 b_2 - h_2 b_1) +
                      Q (g_1 c_2 + g_2 c_1 - h_1 b_2 + h_2 b_1) + \nonumber \\
    &   &             U (-g_1 h_2 + g_2 h_1) +
                      V (g_1 h_1 + g_2 h_2) \nonumber
\end{eqnarray}
where all second order terms in $B$ and $C$ have been omitted.
Somewhat unconventionally I have
defined each complex term above by e.g. $G = g_1 + i g_2$ rather than
$G = g_1 e^{ig_2}$ to simplify the notation. Note
that these parameters are assumed to be time-independent but will
not, in general, be frequency independent.
In a perfect system one has
$b_1=b_2=0$, $c_1=c_2=0$, $g_1=h_1=1$ and $g_2=h_2=0$ and hence the measured
Stokes parameters are identically equal to the true Stokes parameters
(ignoring the effects of rotation of Stokes $Q$ into $U$ along the 
line of sight).

If the gains and instrumental phase have been solved for by some other method,
as is often the case, (e.g. through observations
of a polarized calibration signal injected directly into the feed),
then $g_1=h_1=1$ and $g_2=h_2=0$ and
equation \ref{thelot} simplifies to:
\begin{eqnarray}
\label{simple}
I_m & = & I + U (b_1 - c_1) + V (b_2 + c_2) \\
Q_m & = & Q + U (b_1 + c_1) + V (b_2 - c_2) \nonumber \\
U_m & = & U + I (b_1 - c_1) - Q (b_1 + c_1) \nonumber \\
V_m & = & V + I (b_2 + c_2) - Q (b_2 - c_2) \nonumber
\end{eqnarray}

We also note, as Britton (2000) has done, that a 
relationship holds between the measured and true Stokes parameters via
\begin{equation}
I_m^2 - Q_m^2 - U_m^2 - V_m^2 = K (I^2 - Q^2 - U^2 - V^2)
\end{equation}
where $K$ is time-invariant and is related to the determinant of
the matrix in Equation 11. This expression, the so-called
invariant interval, is not frequency-invariant however, as it depends 
on the (frequency dependent) terms in the $F$ matrix.
Britton's idea of using the invariant interval for timing purposes has
been used to great effect by van Straten et al. (2001).

\section{Effects of feed rotation}
It is sometimes stated that performing a (short) observation, rotating the 
feed through 90\deg and observing again and summing the two observations
has the effect of cancelling the effect of the leakage parameters.
However, this is not the case, even to first order. The effect 
of feed rotation effectively
converts $Q$ to $-Q$ and $U$ to $-U$. To sum the two observations
one essentially averages the sum of the two values of $I_m$ and $V_m$ and 
averages the differences of the two $Q_m$ and $U_m$ measurements.
Assuming the gains are correctly calibrated, then using equation
\ref{simple} above, one obtains:
\begin{eqnarray}
I_m & = & I + V (b_2 + c_2) \\
Q_m & = & Q + U (b_1 + c_1) \nonumber \\
U_m & = & U - Q (b_1 + c_1) \nonumber \\
V_m & = & V + I (b_2 + c_2) \nonumber
\end{eqnarray}
and the total linear polarization is thus
\begin{eqnarray}
L_m & = & L \sqrt{1 + (b_1 + c_1)^2}
\end{eqnarray}
These last two equations show that the measured Stokes parameters 
are now a better approximation
to the true Stokes parameters than a single observation would yield.
However, even though the error in $I_m$ and $L_m$ are likely to be small,
the error in $V_m$ can potentially be large if $V$ is small and $b_2 + c_2$
is significant. The error in the position angle of the linear
polarization will be of order ${\rm tan^{-1}}(b_1 + c_1)$.

A further potentially interesting measurement is to subtract
$V_m$ from one observation with that from an observation where the feed
is rotated through 90\deg. In this case one obtains
\begin{eqnarray}
V_m  & = & -Q (b_2 - c_2)
\end{eqnarray}
If Q is large, one can then obtain $b_2 - c_2$ directly.

\section{Effects of observing a polarized calibration signal}
It is often the case that a cal probe is included in the receiver
package. This probe usually injects a signal
between the two (linear) probes at an angle of 45\deg to both. The cal
thus has $I=U$ and $Q=V=0$. Imagine now that the two probes have
identical gains and there is no instrumental rotation (i.e. we can
set $g_1=h_1=1$ and $g_2=h_2=0$). Then, if the leakage terms
are present, observations of the cal will
yield a signal
\begin{eqnarray}
v_{xx^*} & = & \frac{1}{2} U (1 + 2 b_1) \\
v_{yy^*} & = & \frac{1}{2} U (1 - 2 c_1) \nonumber \\
V_m    & = &             U (b_2 + c_2) \nonumber
\end{eqnarray}
The first two expressions look like gain difference between the
two probes, and the presence of $V$ looks like instrumental rotation
of $U$ into $V$. Hence, even though we started with a system with
perfect gains and phases, we derive using equation \ref{simple}:
\begin{eqnarray}
g_1 & = & \frac{1}{1 + 2 b_1} \\
h_1 & = & \frac{1}{1 - 2 c_1} \nonumber \\
h_2 & = & b_2 + c_2  \nonumber
\end{eqnarray}
and hence the system looks impure!
One then re-observes the cal (or any other pure U signal)
with these terms in equation \ref{thelot} and one obtains:
\begin{eqnarray}
I_m & = & U \left( 1 - \frac{b_1}{(1 + 2 b_1)^2} + \frac{c_1}{(1 + 2 c_1)^2} +
            \frac{1}{2}b_2^2 - \frac{1}{2}c_2^2 \right) \\
Q_m & = & U \left( \frac{-b_1}{(1 + 2 b_1)^2} + \frac{c_1}{(1 + 2 c_1)^2} +
            \frac{1}{2}b_2^2 - \frac{1}{2}c_2^2 \right) \nonumber \\
U_m & = & U \left( \frac{b_1 + 2 b_1^2 - c_1 + 2 c_1^2 + 1}{(1 + 2 b_1)(1 - 2 c_1)} +
            b_2(b_2 + c_2) \right) \nonumber \\
V_m & = & U \left( \frac{2 b_2 (c_1 + b_1)}{(1 + 2 b_1)(1 - 2 c_1)} - 
            b_1(b_2 + c_2) \right) \nonumber
\end{eqnarray}
This implies that one does not, in general, measure a pure $U$ signal
in spite of the calibration procedure.

\section{Circular Feeds}
The Parkes radio telescope has receivers with linear feeds. However, many
other telescopes have circular feed
systems. To apply this method to receivers with circular feeds one needs
to change the Stokes vector $e^S$ to
\begin{equation}
e^S = 
\left( \begin{array}{c}
I \\ V \\ Q \\ U
\end{array} \right)
\end{equation}
and equation 10 to read
\begin{eqnarray}
I_m & = & v_{rr^*} + v_{ll^*} \\
V_m & = & v_{rr^*} - v_{ll^*} \nonumber \\
Q_m & = & v_{rl^*} + v_{lr^*} \nonumber \\
iU_m & = & v_{rl^*} - v_{lr^*} \nonumber  
\end{eqnarray}
The algebra can then be worked through to obtain expressions for the
measured Stokes parameters in terms of the true parameters as
in equation \ref{thelot}.

\section{Application}
From equation \ref{thelot} it can be seen there are 4 measured
quantities (the Stokes parameters) and a total of 12 unknowns.
These 12 unknowns comprise the 4 leakage parameters 
($b_1$, $b_2$, $c_1$, $c_2$), the 4 gain terms ($g_1$, $g_2$, $h_1$, $h_2$),
the total intensity, linear and circular polarization of the source
and an unknown angle $\phi$ which deals with rotation along the line
of sight such that $Q = L {\rm cos}\phi$ and $U = L {\rm sin}\phi$.
This angle is made up of the intrinsic position angle of the source,
the rotation measure through the interstellar medium and ionosphere,
the parallactic angle and the feed angle(s) with respect to some
reference frame. Of these angles, only the ionospheric rotation measure
and the parallactic angle are time variable. Unless some independent
measurement can be made of the ionosphere, this introduces unwanted 
noise into the calibration matrices. Ignoring this,
rotating the feed relative to the sky (for example
as naturally occurs with a feed mounted on an alt-az telescope) can be
used to determine the 12 unknowns. Provided we have sufficient
independent measurements of the Stokes parameters at a variety of
parallactic angles, we can, in principle, determine {\rm all} the
unknowns. Sault et al. (1996) show that observations of an unpolarized
source will yield only 9 of the unknowns, but that observations of
a strongly linearly polarized source can determine all 12.

Pulsars make ideal sources for (self-)calibration as they are generally
highly linearly polarized and have a reasonable degree of circular
polarization. Furthermore, the polarization changes across the pulse
profile and one can thus use the information in multiple phase bins
at a given pointing to help solve for the unknowns.
The main drawback to using pulsars as calibrators is that 
they scintillate which implies that the total intensity, $I$, can vary 
significantly on timescales of order minutes. The fractional polarization
is not affected by scintillation, however, and therefore
to overcome this problem one is forced to normalise the measured Stokes 
parameters by $I_m$ at the expense of being unable to measure the 
true value of $I$. Unfortunately this procedure introduces errors
as, of course, $I_m$ is not equal to $I$. In practice these errors will
be small unless the source is highly polarized and $b_1 - c_1$ is large.

A previous method for calibration of pulsar signals was given in 
the Appendix of Stinebring et al. (1984) and is still in use 
(e.g. Weisberg et al. 1999). However, their method involves a number
of simplifications and assumptions which, in my view, are incorrect
in some cases and which are, in any case,
no longer necessary to make. Any least squares fitting algorithm can
easily solve for 12 unknowns with $\sim$100 data points in a matter
of seconds and the full equations should thus be used.

Data on PSR J1359--6038 were collected on two separate occasions
in July and Novmeber 2000. Each observing session lasted approximately
10 days and we obtained 65 and 66 independent observations covering
the whole parallactic angle range available (approximately $\pm$94\deg).
These data were taken using the 64-m Parkes telescope using the
centre beam of the multibeam
receiver at a central observing frequency of 1318 MHz and 
with a total bandwidth of 128 MHz. At the start of the
observing session observations were made of the flux calibrator
Hydra A to obtain the system equivalent flux density.
Then, a 90 sec observation of the cal signal was made followed immediately 
by a 3 min observation of the pulsar. The pulsar observation was
calibrated for differential gain and phase in the two probes
based on the results from the cal observation and flux
calibrated from the observations of Hydra A. Pulse profiles
were formed in each of $I_m$, $Q_m$, $U_m$ and $V_m$ with
256 phase-bins per profile and 8 frequency channels across
the 128 MHz total bandwidth. $Q_m$, $U_m$ and $V_m$ were then
normalised by $I_m$.

Figure 1 shows the measured (normalised) Stokes parameters as a function of
parallatic angle from data taken in November 2000 from the
peak of the pulse profile in one of the frequency
channels centered at 1365 MHz. It can clearly be seen
that the circular polarization, $V_m$, varies significantly
with parallactic angle and that the variations are in phase with the
variations in $Q_m$. The deviation in $V_m$ is very large with a peak-to-peak
amplitude of about 0.2. The implication of this (from equation \ref{simple})
is that $b_2-c_2$ is of this order, given that $Q_m$ is about unity.
\begin{figure}[h]
\begin{center}
\psfig{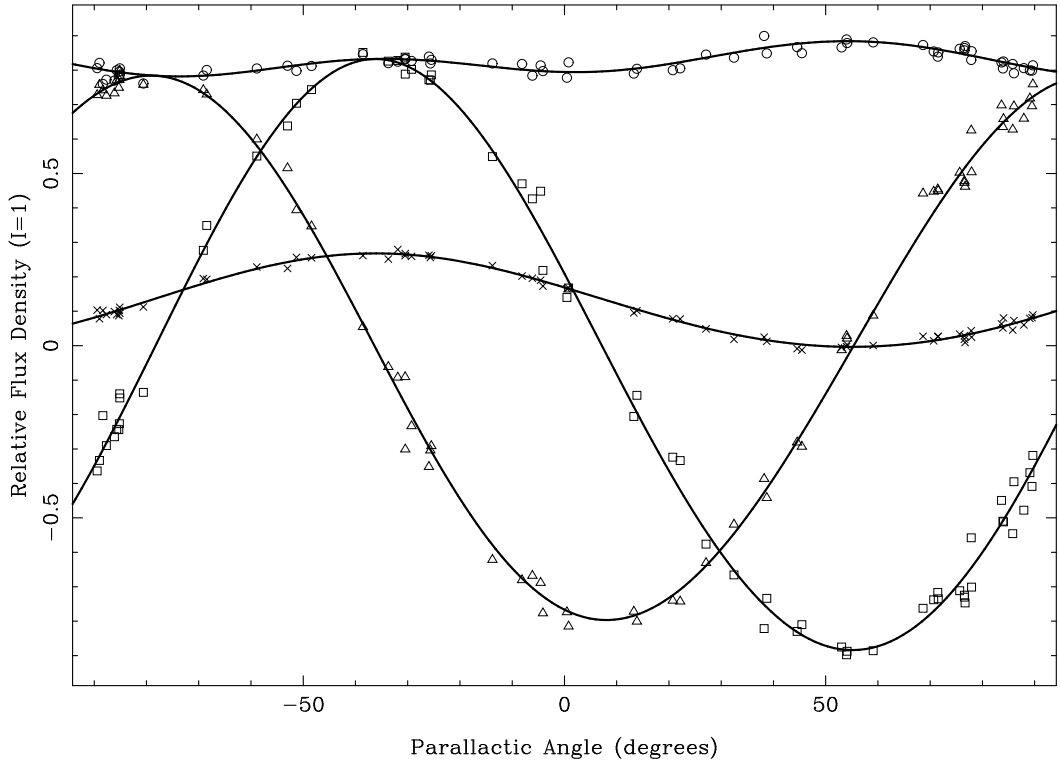}
\caption{Measured Stokes parameters for PSR J1359--6038 as a function
of parallactic angle. Here, $L_m$ is denoted by circles, $Q_m$ by squares,
$U_m$ by triangles and $V_m$ by crosses. The lines denote the best fit
to the data after solving equation \ref{thelot}.}
\end{center}
\end{figure}

It was shown in section 5, that observations of the cal induces 
subsequent errors in the observed Stokes parameters.
These errors can be removed by
(re-)solving for the gains and phase terms (although nominally the cal
observations are used to set these) using equation \ref{thelot}.
Practically, however, I set $g_1 = 1$ and $h_1 = 0$ and leave them fixed;
$g_2$ and $h_2$ then measure the relative gain and phase differences
between the two probes.
I then used a least squares fitting algorithm (in this case the
Levenberg-Maquart Method; LMM)
to minimise the residuals of $(|Q_m-Q| + |U_m-U| + |V_m-V|)$ and
solve for the 9 unknowns from equation \ref{thelot} given the 
data (remembering that $I_m$ is fixed at 1,
all the observed Stokes parameters are normalised by $I_m$
and $g_1=1$, $h_1=0$).
The result of the fitting is shown in the Figure.
At this particular frequency, $b_1=0.0$, $b_2=-0.05$, $c_1=0.02$
and $c_2=0.12$. Given this and the discussion in Section 4, rotating the
feed through 90\deg and observing again still induces a leakage of
I into V of order 7\%, about the same size as the true V signal in
many pulsars. As pointed out, an independent measurement of $b_2 - c_2$
can be obtained in this way, and confirms the measurements derived above.
\begin{figure}[h] \begin{center}
\psfig{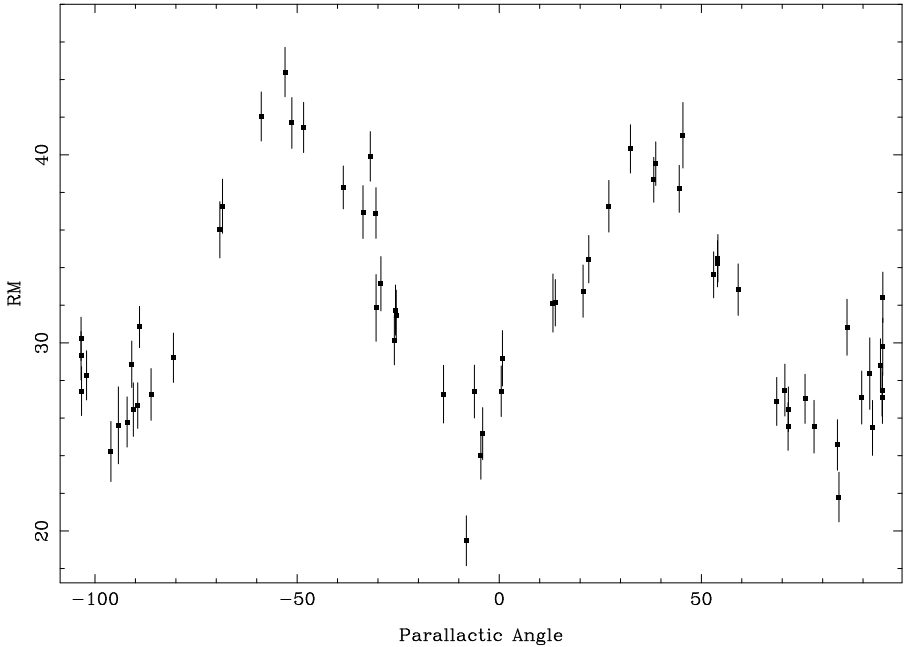}
\caption{Rotation measure of PSR J1359--6038 as a function of
parallactic angle before polarization calibration. The error bars
are formal errors from the fitting process used to determine the RM.}
\end{center}
\end{figure}

At this point, all the parameters of the matrix form of Equation 11
have effectively been determined. The true Stokes parameters can
then be obtained by multiplying the observed Stokes parameters with
the inversion of this matrix. This can be achieved with e.g.
the numerical recipes routine {\sc lubksb}.

Figure 2 shows the rotation measure of the pulsar as a function of 
parallactic angle prior to polarization calibration. There are clear 
systematics in the data with sinusoidal variations 
of $\sim$10 rad~m$^{-2}$ about a mean of 35 rad~m$^{-2}$.
These are due to the frequency dependency of the leakage terms $B$ and $C$.
Figure 3 shows the effects of the calibration. The mean rotation measure 
is now 16 rad~m$^{-2}$ and the systematic effects are largely absent.
\begin{figure}[h]
\begin{center}
\psfig{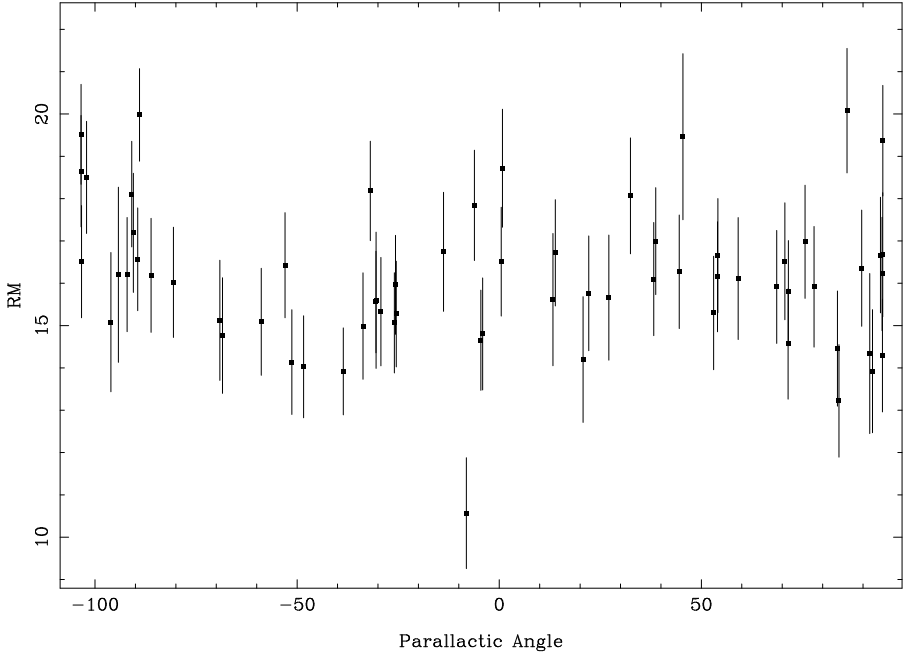}
\caption{Rotation measure of PSR J1359--6038 after polarization
calibration.}
\end{center}
\end{figure}

\section{Conclusions}
Using the excellent Hamaker et al. (1996) paper as a mathematical
foundation, I have derived the matrices linking the true Stokes
parameters to the observed Stokes parameters for single dish observations.
It is then a computationally trivial task to solve for the terms in
this matrix by least squares minimisation and thus provide a 
polarization calibration of the feed and receiver system.

\section*{Acknowledgments}
I would like to thank M. Kesteven for fruitful discussions and a careful
reading of this manuscript and R. Manchester for discussion on
the calibration procedure used at Parkes. The referee, M. Britton,
made a valuable contribution to improving the paper.
A. Karastergiou and L. Nicastro provided assistance with the observations.

\section*{References}
Born, M. \and Wolf, E. 1964, Principles of Optics. Pergamon Press.\\
Britton, M. 2000, ApJ, 532, 1240\\ 
Hamaker, J. P., Bregman, J. D. \and Sault, R. J. 1996, A\&ASS, 117, 137\\
Sault, R. J., Hamaker, J. P. \and Bregman, J. D. 1996, A\&ASS, 117, 149\\
Stinebring, D. R., Cordes, J. M., Rankin, J. M.,
Weisberg, J. M. \and Boriakoff, V. 1984, ApJSS, 55, 247\\
van Straten, W., Bailes, M., Britton, M., Kulkarni, S. R., Anderson, S. B.,
Manchester, R. N., Sarkissian, J. 2001, Nature, 452, 158.\\
Weisberg et al., 1999, ApJSS, 121, 171\\

\end{document}